\documentclass[prb,amsmath,amssymb,showpacs,preprint]{revtex4}
\usepackage{graphicx}
\usepackage{bm}
\def\be{\begin{equation}}
\def\ee{\end{equation}}
\def\A{\tilde A}
\def\TLP{T_{\rm LP}}
\def\ps{\tilde\psi}
\def\L{\tilde L}

\def\R{\tilde R}
\def\nab{\bm{\nabla}}

\begin{document}
\title{Influence of Thermal Fluctuations on Uniform and Nonuniform Superconducting Rings according to the Ginzburg--Landau and the Kramer--Watts-Tobin Models}
\author{Jorge Berger}
\affiliation{Department of Physics and Optical Engineering, Ort Braude College, P. O. Box 78,
21982 Karmiel, Israel}
\email{jorge.berger@braude.ac.il}
\begin{abstract}
We evaluate the influence of thermal fluctuations on superconducting rings that enclose a magnetic flux, using the time-dependent Ginzburg--Landau (TDGL) or the Kramer--Watts-Tobin (KWT) model, while thermal fluctuations are accounted for by means of Langevin terms. This method is applicable in situations where previous methods are not, such as nonuniform loops, rings with large width to radius ratio and loops with large coherence length to perimeter ratio. We evaluate persistent currents, position and statistical behavior of flux-induced vortices and lifetime of metastable fluxoid states.
The influence of nonuniformity on the persistent current does not depend strongly on the details of the cross-section profile; it depends mainly on its first harmonic, but not only on it. As a consequence of nonuniformity the maximum of the persistent current shifts to smaller fluxes and the passage between fluxoid states remains non-hysteretic down to lower temperatures than in the case of a uniform sample. 
Our results using TDGL agree remarkably well with recent measurements of the persistent current in superconducting rings and with measurements of the position of a vortex that mediates between fluxoid states in an asymmetric disk with a hole; they could also provide a plausible explanation for unexpectedly short measured lifetimes of metastable states. Comparison between TDGL and KWT indicates that they lead to the same results for the persistent current, whereas KWT leads to larger lifetimes than TDGL.
\end{abstract}
\pacs{74.40.+k, 05.10.Gg, 73.23.Ra, 74.78.Na}% 
\maketitle

\section{INTRODUCTION}
Superconducting rings have attracted the interest of physicists during many decades, because they are an easily accessible system in which measurable quantities depend on the enclosed flux, rather than on local fields only. More precisely, these quantities depend on the magnetic flux modulo $\Phi_0=hc/2e$, as expected from quantum behavior. One of the milestones in this endeavor was the Little--Parks experiment,\cite{LP} in which the transition temperature is an oscillatory function of the applied flux.

A closer look at the Little--Parks results shows that, as expected from a one-dimensional system, there is actually no phase transition.
The transition region is dominated by thermal fluctuations and what experimentalists call ``transition temperature" is in most cases the temperature at which the resistance becomes some given fraction of the normal state resistance. Since the Little--Parks effect is most pronounced for small samples, fluctuations are especially important. Superconducting rings are a compelling system for studying fluctuations: theoretically, they are a nontrivial system which is not invariant under time reversal, has a rich yet simple phase diagram, and both Gaussian and non-Gaussian fluctuations may be involved; experimentally, the influence of fluctuations is quite directly controllable and detectable.

A theory for the evaluation of fluctuation contribution to conductivity above the critical temperature is due to Aslamazov and Larkin.\cite{AL} An exact method for evaluation of the average current around a uniform loop, as a function of the temperature and enclosed flux, was developed by von Oppen and Riedel (vOR).\cite{Oppen} 
Although the vOR method is exact from the point of view of statistical mechanics (provided that the energy spectrum is described by the Ginzburg--Landau model), it is limited to a static situation and to rings that are perfectly uniform and one-dimensional; moreover, for small values of the parameter $\gamma $ (will be defined in Sec.\ \ref{1Dl}) numeric implementation of this method becomes exponentially difficult.
An independent theory, which discusses the influence of the shape of the loop and uses a two-level approximation is due to Daumens {\it et al.}\cite{Buzdin} Recently, the persistent current in a superconducting loop was analyzed by Schwiete and Oreg.\cite{Oreg} For the regime in which the superconductor--normal transition occurs at finite temperature, they find asymptotic expressions for the current near integer or half-integer flux. For sufficiently small rings, there exists a quantum critical point at which the superconductor--normal transition occurs at zero temperature; in this case they evaluate the persistent current near the critical point by means of the full fluctuation propagator. The current thus obtained is larger than predicted by classical fluctuations alone.

The vOR theory was tested by Zhang and Price.\cite{ZP} They measured the susceptibility and found values larger by an order of magnitude than predicted. Recently, Koshnick {\it et al.}\cite{Nick} repeated this experiment for several samples with widths of the order of 100 nm, using a scanning SQUID, and found good agreement with the theory. Rosario {\it et al.}\cite{Liu} measured conductivity in ultrathin cylinders at half-integer number of fluxoids, and found exponential dependence on the temperature above the transition, in contrast with the power dependence predicted by Aslamazov--Larkin. 

In this article we study the influence of the width and nonuniformity of the loop on the average current and on the transitions between fluxoid states. We focus on the Little--Parks temperature range, i.e.\ the range of temperatures (depending on the flux) for which the superconducting transition would occur in the absence of fluctuations.
%between the temperatures at which in the absence of fluctuations the superconducting transition would occur for integer and for half-integer flux. 
For higher temperatures, currents are small and experimental and numeric values are noisy; for lower temperatures, there is hysteresis and the statistical average becomes irrelevant. Somewhat lower temperatures will also be considered, because fluctuations and nonuniformity lower the temperature at which hysteresis appears and, sporadically, we will enter the hysteretic range. Most of the article is devoted to evaluation of the current around the rings studied in Ref.~\onlinecite{Nick} and comparison with their experimental values. For most samples, our results agree with the vOR theory (and therefore with the theoretical analysis in Ref.~\onlinecite{Nick}); when they do not, we attribute the difference to the width and/or nonuniformity of the samples.

The most widespread theoretical tool in the study of dynamic properties of superconductors (see e.g.\ Refs.\ \onlinecite{AL,McH,AV}) is the time-dependent Ginzburg--Landau model\cite{Sch1} (TDGL). A recent review\cite{AGZ} warns us of the fact that for gapped superconductors TDGL is not strictly valid even above $T_c$, while below $T_c$ it becomes totally wrong; nevertheless TDGL remains popular because of its simplicity and its ability to reproduce observed phenomena. Kramer and Watts-Tobin\cite{KWT} generalized TDGL so that it should be applicable to gapped superconductors as long as there is local equilibrium, while still retaining some of the simplifying features of the TDGL formalism.

The objectives of the present study are (i) to predict the influence of imperfections (i.e., nonuniformity and finite width) on fluctuation superconductivity in rings, (ii) to compare our predictions with experimental results and thus check the range of applicability of TDGL, at least as a phenomenologic model, and (iii) to find how features of fluctuation superconductivity (i.e., average current and lifetime of metastable states) are affected when TDGL is generalized to the Kramer--Watts-Tobin model.

This article is organized as follows. In Sec.~\ref{Meth} we describe our computational method, in Sec.~\ref{Rep} we present general results that stem from our method and in Sec.~\ref{Comp} we use our method to explain measured values. Most of our calculations aim at fitting the currents measured in Ref.~\onlinecite{Nick}, but we also address experiments that measure the position of a vortex\cite{Kanda,JB} and lifetimes of metastable states.\cite{ZP} In each of these sections we distinguish between thin rings for which a one-dimensional treatment should suffice and wide rings for which corrections are necessary. In Sec.~\ref{kwt} we extend our analysis to the Kramer--Watts-Tobin model and the central features of our results are briefly summarized in Sec.~\ref{Disc}. In Appendix \ref{AA} we describe how we dealt with some numerical difficulties and in Appendix \ref{BB} we speculate on the possibility of treating arrays of normal rings by means of the Ginzburg--Landau model.

\section{TDGL with thermal fluctuations\label{Meth}}
The Ginzburg--Landau model describes the behavior of a superconductor by means of an order parameter field $\psi$ and an electromagnetic potential ${\bf A}$. For a thin wire and close to the onset of superconductivity, the magnetic field in the wire can be approximated by the applied field. In this case and in cgs-Kelvin gaussian units the model postulates the energy density $\alpha|\psi |^2+\beta|\psi |^4/2+(\hbar^2/2m)|(i\nab -2\pi{\bf A}/\Phi_0)\psi|^2+{\bf A}\cdot {\bf j}/c$, with $\alpha=12\hbar k_B(T-T_c)/\pi m\ell v_F$, $k_B$ the Boltzmann constant, $T$ the temperature, $T_c$ the critical temperature, $m$ the mass of a Cooper pair, $\ell$ the mean free path, $v_F$ the Fermi velocity, $\beta=8\pi(\kappa \hbar e/mc)^2 $, $\kappa =1.15\times 10^{-6}/\ell$, $-e$ the electron charge and ${\bf j}$ the total current density. The integral of the energy density over the volume of the sample will be denoted by $G$.
All the experiments\cite{ZP,Nick,Kanda} with which we will compare our results were performed on mesoscopic aluminum samples;
 we will therefore adopt $v_F=1.5\times 10^8\,$cm/sec, as appropriate for aluminum with a BCS coherence length 1.6$\mu$m; this value of $v_F$ is smaller than what would be obtained within the free-electron model. We will assume that the samples are in the dirty limit.

The characteristic temperature scale in our analysis will be $T_{\rm LP}=\pi^3\hbar v_F\ell/24k_B L^2$, where $L$ is the perimeter of the ring. $T_{\rm LP}$ is the depression of the transition temperature that would be obtained in the absence of fluctuations for a uniform profile and half-integer magnetic flux (in units of $\Phi_0$). $T_{\rm LP}$ equals the Thouless energy multiplied by $\pi/8k_B$. For the conductivity of normal electrons we have taken the aluminum value $\sigma =1.22\times 10^{11}\ell\Omega^{-1}{\rm cm}^{-2}$, but we have also checked other values.

All the rings that we will consider will be thin and narrow in the scale of the coherence length $\xi=\hbar/\sqrt{2m|\alpha |}$ and the magnetic penetration depth $\kappa\xi$. However, not all the rings will have a width which can be neglected when compared to its average radius; these rings will require a separate treatment.

\subsection{One-Dimensional Loops\label{1Dl}}
A numerical method for the implementation of TDGL with thermal fluctuations was presented in Ref.~\onlinecite{Langevin} for the case of thin wires and loops that do not enclose a magnetic field. In this section we review this reference and point out the required modification when magnetic flux is enclosed.

For computational purposes the loop is divided into $N$ segments of length $L/N$ and the fields and the energy are discretized. ${\bf A}$ can be taken tangential to the wire and we denote by $\psi_k$ and $A_k$ the values of $\psi$ and $A$ in the $k^{\rm th}$ segment. We also define $\A_k =2\pi LA_k/N\Phi_0$. From TDGL without fluctuations it follows that the derivatives of $\psi_k$ and $A_k$ with respect to time have the form
\be
d{\rm Re}[\psi_k]/dt=-\Gamma_{\psi,k} \partial G/\partial {\rm Re}[\psi_k] \;, 
\label{macu}
\ee
with an analogous equation for ${\rm Im}[\psi_k]$, and
\be
d\A_k/dt=-\Gamma_{A,k} \partial G/\partial \A_k \;,
\label{macA}
\ee
where $\Gamma_{\psi,k}=Nmv_F\ell /3\hbar ^2 Lw_k$ and $\Gamma_{A,k}=4e^2L/N\hbar^2\sigma w_k$, with $w_k$ the cross section of segment $k$. 

Equations (\ref{macu}) and (\ref{macA}) have the ``canonic" form $dx/dt=-\Gamma \partial G/\partial x$. In this case the influence of fluctuations is obtained by adding Langevin terms to the right hand side of these equations, such that the integral of a Langevin term over a period of time $\tau$ has gaussian distribution, zero average, and variance $2\Gamma k_BT\tau$.

The electromagnetic potential can be gauged out of the expression for the energy, and therefore from Eq.~(\ref{macu}), by means of the transformation $\ps (s)={\cal U}(s)\psi (s)$ with ${\cal U}(s)=\exp [(2\pi i/\Phi_0)\int_0^s A(s')ds']$. The price of this transformation is that the gauge-invariant order parameter $\ps$ does not obey  a ``canonic" evolution equation. 

Reference \onlinecite{Langevin} adopts several normalizations that are appropriate for the fluctuation region $T\approx T_c$. The typical size of the order parameter is $\bar\psi=(2mk_B^2T_c^2/\beta \hbar^2\bar w^2)^{1/6}$, where $\bar w$ is the average cross section, the typical coherence length is $\xi_\beta=(\bar w\Phi_0^2/32\pi^3\kappa^2k_BT_c)^{1/3}$ and the time unit is $\bar\tau=3\xi_\beta^2/v_F\ell$. Under free electron gas assumptions, the parameter $\gamma $ of Ref.~\onlinecite{Nick} can be expressed as $\gamma =0.5(L/\xi_\beta)^3$. We then define the normalized quantities $\L=L(2mk_BT_c)^{1/2}/\hbar$, $\R=6e^2 L/\sigma m\bar w v_F\ell$, $\psi_{\beta k}=\psi_k/\bar\psi$, $\ps_{\beta k}=\ps_k/\bar\psi$ and $\alpha '=\alpha /\beta\bar\psi^2$. For the aluminum values and the units that we have adopted, $\xi_\beta =0.679(\bar w/\kappa^2 T_c)^{1/3}$, $\tilde{L}=6.72\times 10^5 LT_c^{1/2}$, $\tilde{R}=4.62\times 10^{-23}L/\ell^2\bar w$, $\alpha '=8.55\times 10^3 \xi_\beta^2 (T-T_c)/\ell T_c$ and $\bar\tau =2\times 10^{-8}\xi_\beta^2/\ell$.
With these normalizations, Eqs. (\ref{macu}) and (\ref{macA}) become
\begin{eqnarray}
\Delta_{\rm mac}\ps_{\beta k}&=&-(\tau /\bar\tau ) [(\alpha '+|\ps_{\beta k}|^2)\ps_{\beta k}+
(N\xi_\beta/L)^2 (2w_k)^{-1} \nonumber \\
&&[(w_k+w_{k+1})(\ps_{\beta k}-\ps_{\beta k+1})+(w_k+w_{k-1})(\ps_{\beta k}-\ps_{\beta k-1})]]\;, \label{Dpsibeta}\\
\Delta_{\rm mac}\A_k&=&-(C_A\tau /w_k)[I\hbar /ek_BT+
(\xi_\beta/L){\rm Im}[\delta_k+\delta_{k+1}]] \;, \label{DAbeta}
\end{eqnarray}
where $C_A=(\bar w\R/2N\bar\tau )(\xi_\beta\L/L)^2$, $I$ is the current around the ring, $\delta_k=N(w_{k-1}+w_k)\ps_{\beta k-1}^*\ps_{\beta k}/\bar w$, $^*$ denotes complex conjugation, $\delta_{N+1}=\delta_{1}$, $w_0=w_N$ and $\ps_{\beta 0}=\exp (-2\pi i\Phi/\Phi_0)\ps_{\beta N}$, where $\Phi$ is the flux enclosed by the ring. $\Delta_{\rm mac}$ denotes the macroscopic (i.e., it ignores fluctuations) increment during a short period of time $\tau$.

The fluctuations in the increment of $\A_k$ have a variance $4C_A\tau /w_k$; the fluctuations of $\ps_{\beta k}$ are taken into account by adding to the real and to the imaginary part terms with variance $N\tau\bar w\xi_\beta /\bar\tau w_k L$ and then modifying its phase in order to take into account the fluctuation in ${\cal U}$.

In this study we assume that nonuniformity enters the problem through the cross section of the ring; we expect that chemical nonuniformity would have a similar effect.

Since the induced flux is negligible, if the applied flux enclosed by a ring remains constant, then $\sum \A_k$ is also constant. This condition determines the current $I$; from Eq.~(\ref{DAbeta}) it leads to
\be
I=\frac{ek_BT}{\hbar}\frac{\sum\eta_k/\tau-(C_A\xi_\beta /L){\rm Im}[\sum w_k^{-1}(\delta_k+\delta_{k+1})]}{C_A\,\sum w_k^{-1}}\;,
\label{condI}
\ee
where the sums are over the $N$ segments of the ring and $\eta_k$ is the contribution of the Langevin term to $\A_k$ during the period of time $\tau$.

Quantities of interest are evaluated by explicit Euler--Mayurama steps following Eqs. (\ref{Dpsibeta}) and (\ref{DAbeta}). We start from an arbitrary set of values for the order parameter, the initial steps are intended for relaxation to some typical state, and subsequent steps are used for statistical averaging.

The only modification that is required in order to apply the procedure of Ref.~\onlinecite{Langevin} is in the boundary condition for $\ps$, which in the case that flux $\Phi$ is enclosed becomes $\ps (L)=\exp (2\pi i\Phi /\Phi_0)\ps (0)$.

\subsection{Wide Rings\label{widesecII}}
Although theory is ``cleaner" for 1D loops, in real life this idealization might not be justified. Moreover, wide rings carry larger currents and may provide us with reliable measurements in fluctuation regions of the $\Phi -T$ plane, where thin rings have a low signal to noise ratio. It will therefore prove useful to develop a simple model for rings that are not ideally 1D.

We consider superconducting rings with a width that is very small compared with the lengths over which the superconducting variables change significantly (coherence length, effective magnetic penetration depth), but is not sufficiently small compared with the radius. As a consequence, the fluxes enclosed by the inner and by the outer boundaries will not be the same. Nevertheless, we will build a quasi-1D model in which the superconducting variables will be represented by quantities that do not depend on the radial coordinate.

Let us denote by $R$ the average radius and the width by $D(\theta )$; the thickness is $w(\theta )/D(\theta )$. 
The ring occupies the region $R-D/2\le r\le R+D/2$ and is immersed in a uniform perpendicular magnetic field $B$. At this stage we neglect the induced magnetic field and $B$ will be taken as constant. 

With the purpose of averaging over $r$, we separate the electromagnetic vector potential into a contribution of the applied magnetic field and a contribution $a$ due to thermal fluctuations, i.e. we write  as ${\bf A}=(Br/2+a)\hat\theta$. In order to have a 1D model we assume that $\psi$ and the voltage are functions of $\theta$ and time only, which implies
\be
a=a_R \frac{R}{r}, \;\;\;\;\; \nab\psi=\psi'_R \frac{R}{r} \hat\theta \;,
\label{rad}
\ee
where $a_R$ is the value of $a$ at $r=R$ and $\psi'_R$ is the derivative of $\psi$ with respect to the arclength $s=R\theta$, taken along the circle $r=R$. 
We will also approximate the current density by the form ${\bf j}=j_R (R/r)\hat\theta$, with $j_R$ independent of $r$ and proportional to $1/w(\theta )$.

We now integrate the energy density over $R-D/2\le r\le R+D/2$ and neglect terms of order higher than $(D/R)^2$. In this approximation the volume averages are $\langle (r/R)^n\rangle =1+n(n+1)(D/R)^2/24$ and the free energy becomes
\begin{eqnarray}
G&=&\oint wds\{\alpha|\psi |^2+\beta|\psi |^4/2+(\hbar^2/2m)[(2\pi/\Phi_0)^2|\psi |^2(B^2(4R^2+D^2)/16 \nonumber \\
&&+a_R BR+a_R^2f_D)+|\psi'_R|^2f_D \nonumber \\
&& +(2\pi i/\Phi_0)(BR/2+a_R f_D)(\psi {\psi'}_R^*-\psi^* \psi'_R)]+(BR/2+f_Da_R)j_R/c\} \;,
\label{Gds}
\end{eqnarray}
where $f_D=1+D^2/12R^2+O((D/R)^4)$.

We can now introduce a new variable, $\hat A =(2\pi /\Phi_0)(BR/2f_D+a_R)$, and rewrite $G$ in the form
\begin{eqnarray}
G&=&\oint wds\{[\alpha+(eBD)^2/(6mc^2)]|\psi |^2+\beta|\psi |^4/2+ \nonumber \\
&& (f_D\hbar^2/2m)[|\psi '_R|^2+\hat A^2|\psi |^2+i\hat A(\psi {\psi'}_R^*-\psi^* \psi'_R)]+f_D\Phi_0\hat Aj_R/2\pi c\} \;.
\label{Gdskova}
\end{eqnarray}
Discrete variables can now be defined as in Ref.~\onlinecite{Langevin}: $\psi_k$ as the average of $\psi$ in cell $k$ and $\tilde A_k$ as the integral of $\hat A$ along the length $L/N$ of this cell. It can be checked that these variables behave as ``canonic" in the sense of Ref.~\onlinecite{Langevin}. 
Integrating over $r$ the TDGL expression for $d\psi /dt$, under natural averaging assumptions we recover Eq.~(\ref{macu}), with $G$ given by Eq.~(\ref{Gdskova}). However, using $(\Phi_0/2\pi)d\hat A/dt=da_R/dt=\langle dA/dt\rangle=-c\langle j_N\rangle /\sigma$, where $j_N$ is the normal current density, we obtain that the coefficient $\Gamma_{A,k}$ is smaller by a factor $f_D$ than the expression obtained for a 1D ring. The same simplifications that were obtained in Ref.~\onlinecite{Langevin} for 1D loops can be achieved here by defining the gauge-invariant order parameter $\ps (s)=\hat{\cal U}(s)\psi (s)$ with $\hat{\cal U}(s)=\exp\left(i\int_0^s \hat A(s')ds'\right)$.

By detailed comparison we conclude that a ring with non-negligible width can be treated by means of Eqs. (\ref{Dpsibeta}) and (\ref{DAbeta}) and the fluctuations described under them, provided that we include the following corrections: (i) $\alpha$ has to be replaced with $\alpha+(eBD)^2/(6mc^2)$, which means adding $(1/3)(\Phi D\xi_\beta /\Phi_0R^2)^2$ to $\alpha '$; (ii) the term proportional to $\xi_\beta^2$ in Eq.~(\ref{Dpsibeta}) has to be multiplied by $f_D$ and (iii) the variance of $\tilde A_k$ has to be divided by a factor $f_D$.

As long as we neglect the self inductance, the integral of the electric field (and hence of $da_R/dt$) around the ring has to vanish; in this case the instantaneous current is given by Eq.~(\ref{condI}).
If there is non-negligible influence of the ring self inductance, which we denote by $L_s$, the total flux through the ring will not be the applied flux $\Phi_x$, but rather $\Phi=\Phi_x+cL_sI$. This has two consequences. The first is that $\oint {\bf A}\cdot d{\bf s}$ varies with $I$ and Eq.~(\ref{condI}) generalizes to
\be
I=\frac{(2eL_s/\hbar\tau)I_{\rm prev}+\sum\eta_k/\tau-(C_A\xi_\beta /L){\rm Im}[\sum w_k^{-1}(\delta_k+\delta_{k+1})]}
{(2eL_s/\hbar\tau)+(\hbar C_A/ek_BT)\sum w_k^{-1}}\;,
\label{condLs}
\ee
where $I_{\rm prev}(t)=I(t-\tau)$.
For the range of parameters that we considered, this refinement turned out to have no noticeable effect. The second consequence is that when comparing our results with experiments, we should present the results as functions of $\Phi_x=\Phi-cL_sI$, since usually this is the controlled quantity. For the parameters we considered, $|\Phi_x-\Phi|$ is small but noticeable.

%We can also use the gauge-invariant method of Sec. II D of Ref.~\onlinecite{Langevin} by defining $\ps (s)=\hat{\cal U}(s)\psi (s)$ with
%$
%\hat{\cal U}(s)=\exp\left(i\int_0^s \hat A(s')ds'\right)
%$.
%From this definition it follows that
%$
%\psi'_R=-i\hat A \psi+\hat{\cal U}^*\ps '_R \;.
%$
%Introducing this expression into Eq.~(\ref{Gdskova}) we obtain
%\be
%G=\oint wds\left\{\left[\alpha+\frac{(eBD)^2}{6mc^2}\right]|\ps |^2+\frac{\beta}{2}|\ps |^4+\frac{f_D\hbar^2}{2m} |\ps'_R|^2  +\frac{f_D\Phi_0\hat Aj_R}{2\pi c}\right\} \;.
%\label{Gaugeds}
%\ee

\section{Representative Results\label{Rep}}
\subsection{One-Dimensional Loops}
Our formalism enables us to study samples with cross section given by any periodic function $w(\theta)$ of the angular polar coordinate by $\theta$. Therefore, our first question is what nonuniformity profiles would it be interesting to study. Since any periodic function can be written as a Fourier series $w(\theta)=\bar w(1+\sum\beta_j\cos j\theta+\sum\gamma_j\sin j\theta)$, another way to pose the question is what Fourier coefficients $\beta_j$ and $\gamma_j$ would it be interesting to consider.

We will restrict ourselves to rings with mirror symmetry $w(-\theta)=w(\theta)$, so that $\gamma_j=0$, and will have to decide on the choice of the $\beta_j$'s. We may be guided by our results for rings at $T\approx T_c$ and small deviations from uniformity, in the absence of fluctuations.\cite{JB,book,JB0} In those studies, which were corroborated by other groups,\cite{Baelus,jumps} it was found that nonuniformity can qualitatively modify the Little--Parks phase diagram and there is a range of temperatures for which there is a continuous passage between consecutive fluxoid states. At the lower end of this range, there is a critical point P$_2$ such that below P$_2$ the passage between fluxoid states is hysteretic and such that at P$_2$ the derivative of the current with respect to the flux diverges. The important aspect of our previous results for the question at hand is that the salient features of the phase diagram, such as the position of P$_2$, depend only on the eccentricity $\beta_1$, whereas higher harmonics have no influence at the leading order. It is therefore natural to consider a shape in which only the first harmonic is present, i.e. $w(\theta)=\bar w(1+\beta_1\cos\theta)$. This profile will be called ``sinusoidal."

\begin{figure}
\scalebox{0.85}{\includegraphics{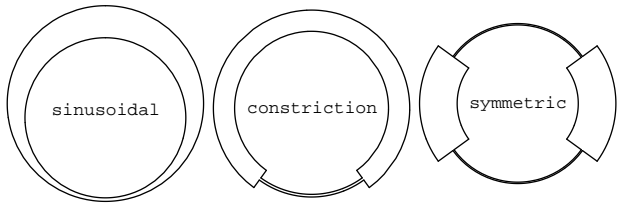}}%
\caption{\label{shapes}Cross section profiles considered in this study.}
\end{figure}

In order to gain some understanding of how other profiles behave, we consider cross sections with discontinuous functions $w(\theta)$, since in this case the coefficients $\beta_j$ have the slowest decrease with $j$, providing a complementary situation to the sinusoidal profile. The simplest discontinuous profile is that of a ring with a constriction. This situation might also be of experimental interest, since the constriction might represent a Josephson junction or a defect. Finally, in order to have an opposite case to the sinusoidal profile, we consider the case of rings with an additional mirror symmetry $w(\pi-\theta)=w(\theta)$, since in this case $\beta_1=0$. A profile like this will be called ``symmetric." Figure \ref{shapes} shows the nonuniformity profiles which we have considered.

\begin{figure}
\scalebox{0.85}{\includegraphics{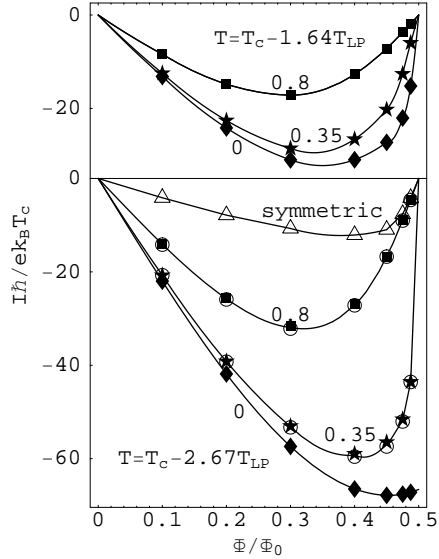}}%
\caption{\label{consT}Average current as a function of the magnetic flux for fixed temperatures and various cross section profiles. In all cases we took $L=2\pi\times 10^{-4}$cm, $\bar w=6.38\times 10^{-11}$cm$^2$, $T_c=1.251$K, $\ell =2.66\times 10^{-6}$cm, 10 computational cells, $2\times 10^8$ averaging steps, $6\times 10^7$ relaxation steps, each step of 10$^{-6}$ time units. Meaning of the symbols: $\blacklozenge$ uniform cross section; $\bigstar$ sinusoidal with $\beta_1=0.35$; $\blacksquare$ sinusoidal with $\beta_1=0.8$; $\bigcirc$ profile with constriction; $\triangle$ symmetric profile. For the profiles with constriction the cross sections of the narrow and the wide parts are in the ratios 1:2.14 and 1:7.97; for the symmetric profile they are in the ratio 1:27.3. The curves are guides for the eye.} 
\end{figure}

Figure~\ref{consT} shows the current $I$ as a function of the magnetic flux $\Phi$ for fixed temperatures. The choice of geometric and material parameters was inspired by those of the samples for which data are available.\cite{Nick} The upper panel is for $T=T_c-1.64T_{\rm LP}$ and all the curves represent sinusoidal cross section profiles. We see that nonuniformity has two qualitative effects: one of them is overall decrease of the current and the second is a shift of the maximum current to lower fluxes, so that the decay of $I$ as $\Phi \rightarrow \Phi_0/2$ becomes more gradual. In contrast to the well known behavior in the absence of thermal fluctuations, even for uniform cross sections and $T<T_c-\TLP$, the average current vanishes at $\Phi =\Phi_0/2$  and the passage between consecutive winding numbers is smooth.

The lower panel in Fig.~\ref{consT} is for $T=T_c-2.67T_{\rm LP}$. For a uniform sample at this temperature, the average current does not go to zero for $\Phi \rightarrow \Phi_0/2$. According to a blind application of statistical mechanics, this average should vanish, since the energies for clockwise or anti-clockwise currents are equal; however, there is an energy barrier between the two fluxoid states, so that in real experiments or simulations only one of them is probed. For all the nonuniform samples shown in the graph, the current does go to zero as $\Phi \rightarrow \Phi_0/2$, meaning that either there is no energy barrier, as is the case for temperature above that of P$_2$, or the barrier is not large in comparison to thermal fluctuations. 

The profiles with a constriction in Fig.~\ref{consT} had a segment of length $0.8L$ with large cross section and a constriction of length $0.2L$. We see that the $I(\Phi )$ curves for these profiles practically coincide with those of appropriate sinusoidal profiles. (Only one line was drawn in these cases for both sets of symbols.) From here we may adopt the working assumption that for any reasonable profile there will be an equivalent sinusoidal profile, so that by studing sinusoidal profiles we may expect to obtain most of the interesting information. Defining the eccentricity as the first harmonic of the cross section divided by its average, we obtain that the constricted profile equivalent to $\beta_1=0.35$ has eccentricity 0.22 and the sample equivalent to $\beta_1=0.8$ has eccentricity 0.4. It follows that the $I(\Phi )$ curve is influenced most significanly by the eccentricity, but not only by it. As an extreme example, we considered a symmetric sample for which the eccentricity is zero. This sample consists of two wide segments, each of length $0.2L$, connected by constrictions of length $0.3L$. In spite of the very large ratio between the cross sections, which manifests itself in a strong inhibition of the average current, the position of the maximum remains quite close to $\Phi = \Phi_0/2$. As a general observation we might say that the inluence of nonuniformity on $I(\Phi )$ is stronger for lower temperatures and close to $\Phi_0/2$.

\begin{figure}
\scalebox{0.85}{\includegraphics{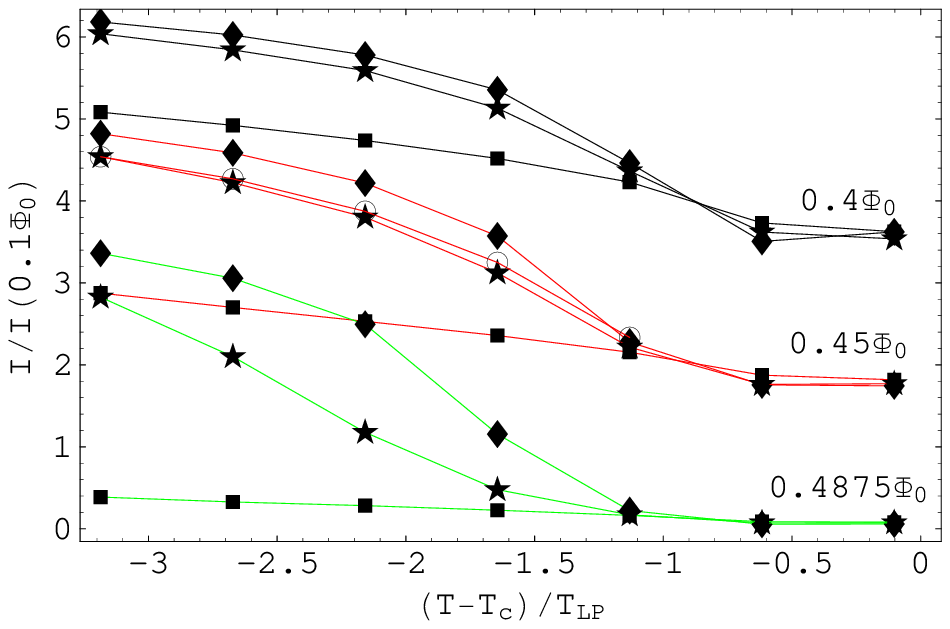}}%
\caption{\label{fixedfi}Current ratios $I(0.4\Phi_0)/I(0.1\Phi_0)$, $I(0.45\Phi_0)/I(0.1\Phi_0)$ and $I(0.4875\Phi_0)/I(0.1\Phi_0)$ as functions of the temperature, for nonuniformities given by $\beta_1=$0, 0.35 and 0.8. The parameters and symbol meanings are the same as in Fig.~\ref{consT}. The circles are for a sample with a constriction of length $0.2L$ and cross section equal to 1/2.14 of that of the rest of the sample; they are shown for $I(0.45\Phi_0)/I(0.1\Phi_0)$ only. The calculated symbols have been joined by straight lines. For visibility, the curves for $I(0.45\Phi_0)/I(0.1\Phi_0)$ (red online) have been raised by 1.5 units and those for $I(0.4\Phi_0)/I(0.1\Phi_0)$ (black) have been raised by 3 units.} 
\end{figure}

Figure \ref{fixedfi} shows the ratio of the current at a given fixed flux to the current at flux $\Phi =0.1\Phi_0$, as a function of temperature. Again, we see that the influence of nonuniformity increases as $\Phi$ approaches $\Phi_0/2$ and, for the parameters used in this graph, is noticeable for $T\alt T_c-1.1T_{\rm LP}$.

\subsection{Wide Rings\label{widesec}}
\begin{figure}
\scalebox{0.85}{\includegraphics{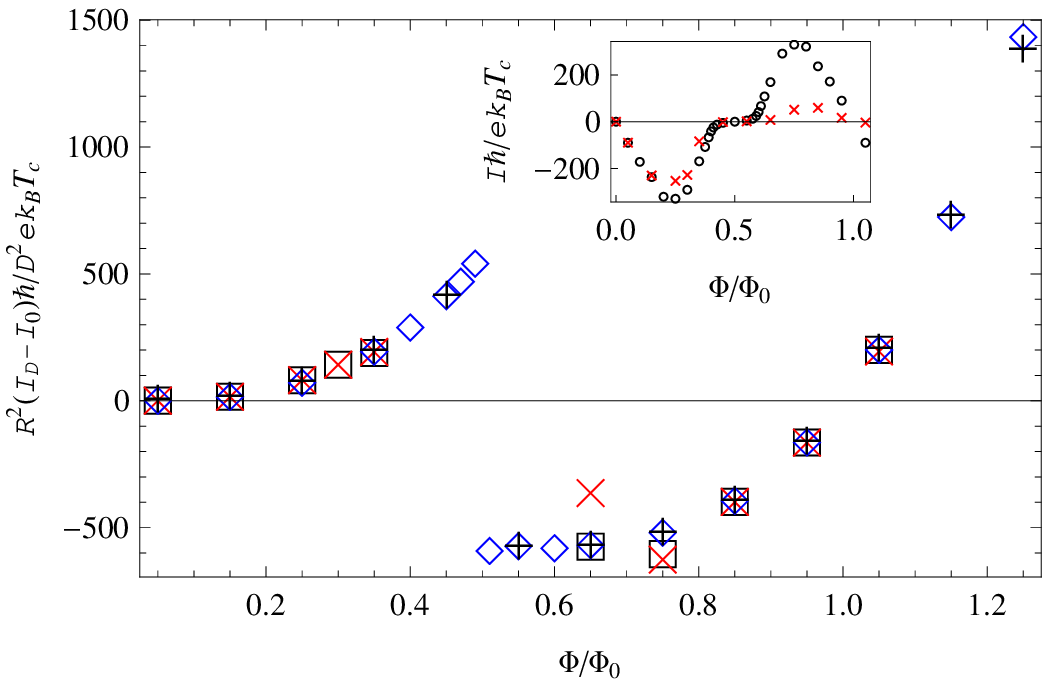}}%
\caption{\label{noflucreg}Reduction of the current as a function of the magnetic flux, outside the fluctuation regions. Calculations are for a sample with $R=3.5\times 10^{-5}$cm, $\bar w=7.83\times 10^{-11}$cm$^2$, $T_c=1.24$K, $\ell =3.02\times 10^{-6}$cm; computational parameters as in Fig.~\ref{consT}. Symbols: $\square$ $T=T_c-0.67\TLP$, $D/R=1/3$; $\times$ (red online) $T=T_c-0.67\TLP$, $D/R=2/3$; $\Diamond$ (blue online) $T=T_c-1.17\TLP$, $D/R=1/3$; $+$ $T=T_c-1.17\TLP$, $D/R=2/3$. The value for $T=T_c-0.67\TLP$, $D/R=2/3$ and $\Phi=0.65\Phi_0$ lies within the fluctuation region $0.38\le \Phi/\Phi_0\le 0.70$ and has been included for comparison. Inset: currents for 1D and quasi-1D samples as functions of the flux; $\circ$ $T=T_c-0.67\TLP$, $D=0$; $\times$ (red online) $T=T_c-0.67\TLP$, $D/R=2/3$.}
\end{figure}

We consider first the case in which the cross section is uniform along the ring. We denote by $I_D$ the current around the ring for non-negligible width, and by $I_0$ the current in the case $D\rightarrow 0$, $w/D\rightarrow\infty$ (the cross section is the same in both cases). In all the cases we considered, we found $0<I_D/I_0<1$, i.e., the finite width inhibits the current, but does not wipe it out. 
The inset in Fig.~\ref{noflucreg} compares $I_D$ and $I_0$ for a typical situation.

In order to make more quantitative statements, we classify our results into those outside and within the fluctuation regions. In the absence of fluctuations, our quasi-1D model leads to vanishing order parameter when
\be
(1/3)(D\Phi/R\Phi_0)^2+f_D(\Phi/\Phi_0-n)^2=(T_c-T)/4\TLP \;,
\label{Phinormal}
\ee
where $n$ is the winding number.
Let us denote by $\Phi_{n-}$ ($\Phi_{n+}$) the smaller (larger) root of Eq.~(\ref{Phinormal}); the regions $\Phi_{n+}<\Phi <\Phi_{(n+1)-}$, where the sample would be normal in the absence of fluctuations, will be called ``fluctuation regions."

Figure~\ref{noflucreg} shows the difference $I_D-I_0$ outside the fluctuation regions for two temperatures and two widths, normalized by the factor $(D/R)^2$. The four sets of points lie in a nearly universal curve, indicating that for the tested range $I_D-I_0$ is proportional to $D^2$ and its value is independent of $T_c-T$.

%\begin{figure}
%\scalebox{0.85}{\includegraphics{inhibit.eps}}%
%\caption{\label{inhibit}Reduction of the current as a function of the magnetic flux. As a guide to the eye, the calculated points have been joined. The %upper lines (red online) are for $T=T_c-1.17\TLP$ and the lower lines (black) for $T=T_c-0.67\TLP$.
%$R=3.5\times 10^{-5}$cm, $\bar w=7.83\times 10^{-11}$cm$^2$, $T_c=1.24$K, $\ell =3.02\times 10^{-6}$cm; computational parameters as in Fig.~\ref{consT}. %$\square$ $D/R=1/3$; $\times$ $D/R=2/3$. The inset shows the fluctuation tail for $D\rightarrow 0$ and $T=T_c-0.67\TLP$.}
%\end{figure}

\begin{figure}
\scalebox{0.85}{\includegraphics{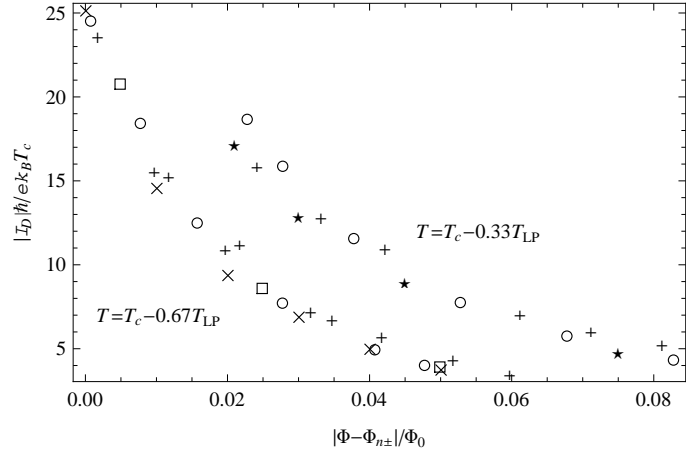}}%
\caption{\label{fluctregion} Current within fluctuation regions in the vicinity of the region border, $\Phi_{n\pm}$, for two fixed temperatures. For visibility, the values for $T=T_c-0.33\TLP$ have been shifted 0.02 units to the right. Symbols: $\square$ $\Phi_{n\pm}=\Phi_{0+}$, $D/R=1/3$; $\times$ $\Phi_{n\pm}=\Phi_{0+}$, $D/R=2/3$; $+$ $\Phi_{n\pm}=\Phi_{1-}$, $D/R=1/3$; $\star$ $\Phi_{n\pm}=\Phi_{1+}$, $D/R=1/5$; $\circ$ $D=0$, arbitrary $\Phi_{n\pm}$. Other parameters as in Fig.~\ref{noflucreg}.}
\end{figure}

Figure~\ref{fluctregion} shows the current within fluctuation regions for the range $|\Phi-\Phi_{n\pm}|\alt 0.06\Phi_0$. If the flux is measured from $\Phi_{n\pm}$, $I_D(\Phi)$ is almost independent of $D$ (but not of the temperature) for $D/R\alt 0.7$ (respectively 0.3, 0.2) in the case $\Phi_{n\pm}=\Phi_{0+}$ (respectively $\Phi_{1-}$, $\Phi_{1+}$).

\begin{figure}
\scalebox{0.85}{\includegraphics{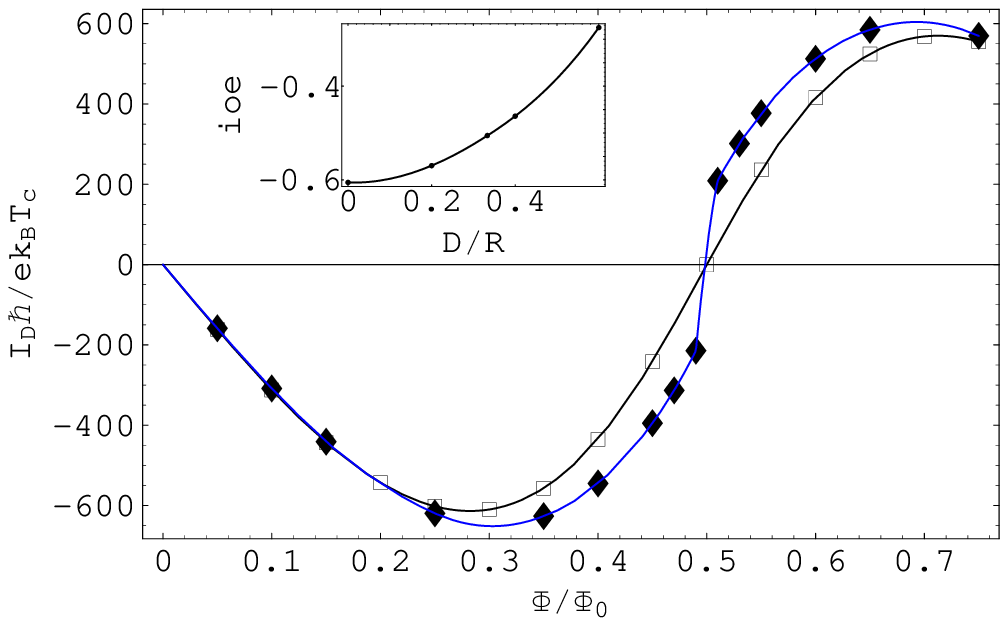}}%
\caption{\label{ioe} Current as a function of the flux for samples with width $D=R/3$ at $T=T_c-1.17\TLP$. The curves are guides to the eye. The rhombs (joined by a blue online curve) are for a uniform sample, and the squares are for $w(\theta)=\bar w(1+0.7\cos\theta)$. The other parameters are as in Fig.~\ref{noflucreg}. In order to compare the shapes, the results for $\beta_1=0.7$ have been multiplied by 1.37. The inset shows an index that represents the influence of eccentricity on the shape of the current, as a function of the width of the sample. ``ioe" stands for $I_D(0.7,0.45)/I_D(0.7,0.1)-I_D(0,0.45)/I_D(0,0.1)$.}
\end{figure}

Let us now consider nonuniform cross sections. We have to deal separately with the case in which nonuniformity is due to the thickness, so that $D$ is constant, and the case in which nonuniformity is due to the width, so that $D$ is a function of $\theta$. For the parameters we considered, there was no significant difference; the results we report in the following are for uniform $D$. Figure \ref{ioe} compares the shapes of the current-flux curve for a uniform and a nonuniform sample. As in the case of 1D loops, nonuniformity brings about a more gradual decay at $\Phi\sim 0.5\Phi_0$.

We would finally like to know whether the influence of eccentricity is enhanced or weakened by the width of the ring. Denoting by $I_D(\beta_1,\phi)$ the current for a ring of width $D$ with eccentricity $\beta_1$ at flux $\phi\Phi_0$, we may regard the ratio $I_D(\beta_1,0.45)/I_D(\beta_1,0.1)$ as a representative index for the shape of the function $I_D(\beta_1,\phi)$, and the difference $I_D(0.7,0.45)/I_D(0.7,0.1)-I_D(0,0.45)/I_D(0,0.1)$ as a representative index of the influence of eccentricity on the shape of $I_D(\beta_1,\phi)$. We see from the inset in Fig.~\ref{ioe} that for wider samples the influence of nonuniformity is smaller.

In our calculations we used several values for the conductivity, with a ratio $\sim 3\times 10^2$ between the largest and the smallest. The value of $\sigma $ had no appreciable influence on the average current. This can be understood, since the average current is an equilibrium quantity and $\sigma $ only affects the dynamics.

\section{Comparison with Experiments\label{Comp}}
\subsection{One-Dimensional Loops}
We have used our method to analyze the 15 samples studied in Ref.~\onlinecite{Nick}.\cite{data} Samples for which $D<R/4$ were treated as ``one-dimensional."
We modeled each sample by a ring with sinusoidal cross section profile. For a given temperature, our first task was the generation of a representative curve $I(\Phi)$ for the experimental data. As the first step we subtracted the linear background; in some cases, we assumed that for a 1D sample the maximum value of $|I|$ in the range $0<|\Phi|<0.5\Phi_0$ has to be the same as for $0.5\Phi_0<|\Phi|<\Phi_0$. 
We then shifted the origin of the applied field to inforce $\int_{-\Phi_0/2}^{\Phi_0/2} I(\Phi)\cos(2\pi\Phi/\Phi_0)d\Phi=0$, and finally fitted the data in the range $-0.5\Phi_0\le\Phi\le 0.5\Phi_0$ to a high order odd polynomial that vanishes at $\Phi=\pm 0.5\Phi_0$. 

The second task was the choice of parameters $T_c$, $\ell$, $\beta_1$ and the mutual inductance between the sample and the scanning SQUID. $T_c$ was taken from Ref.~\onlinecite{Nick}. $\ell$ was chosen by assuming a uniform ring, considering the highest temperature for which measurements are reasonably reproducible and periodic, and then requiring that our calulated values reproduce the experimental smoothed ratio $I(\Phi_1)/I(\Phi_2)$ where, typically, $\Phi_1\sim 0.1\Phi_0$ and $\Phi_2\sim 0.35\Phi_0$. In order to fix $\beta_1$, we again require agreement with the experimental ratio $I(\Phi_1)/I(\Phi_2)$, but this time we consider the lowest temperature for which there is no hysteresis and $\Phi_2\sim 0.47\Phi_0$. After $\beta_1$ was fixed, we re-evaluated $\ell$, using the obtained value of $\beta_1$ rather than $\beta_1=0$; we found that this second iteration did not modify $\ell$ significantly. Finally, the mutual inductance was fixed by fitting the calulated points to the experimental $I(\Phi)$ for the lowest temperature and in the range $0<\Phi\le 0.4\Phi_0$. The mutual inductances we found are typically $(10\pm 5)\%$ smaller than those reported in Ref.~\onlinecite{Nick}. This small discrepancy can be attributed to the experimental uncertainty in the sample-SQUID distance.

\begin{figure}
\scalebox{0.85}{\includegraphics{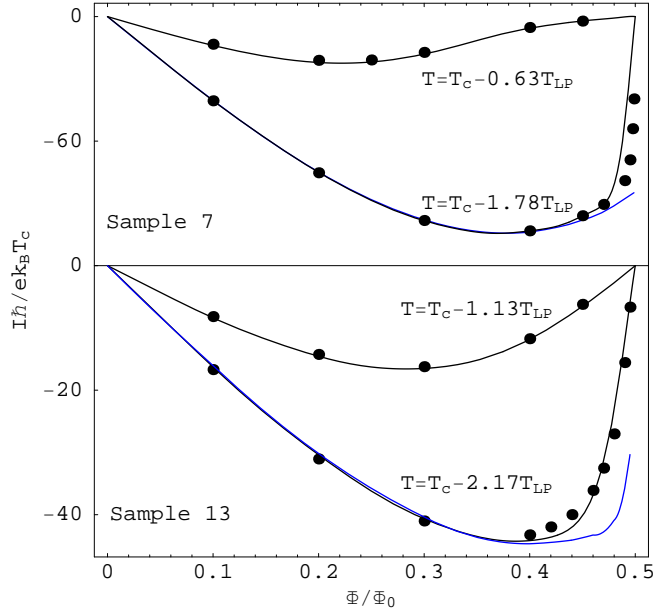}}%
\caption{\label{Nicdata}Current as a function of the magnetic flux for two samples that appear to be nonuniform. The lines marked by temperatures are experimental and the points were calculated. The lowest line (blue online) was calculated for $\beta_1=0$. Parameters for sample 7 taken from Ref.~\onlinecite{Nick}: $L=3.14\mu$m, $\bar w=3.8\times 10^{-11}{\rm cm}^2$, $T_c=1.264$K; parameters for sample 7 fitted here: $\ell=20.5$nm, $\beta_1=0.13$ and mutual inductance $M_{S-R}=0.079\mu\Phi_0/$nA. For the line $\beta_1=0$ we took $M_{S-R}=0.078\mu\Phi_0/$nA. For sample 13, $L=6.28\mu$m, $\bar w=6.4\times 10^{-11}{\rm cm}^2$, $T_c=1.251$K, $\ell=26.6$nm, $\beta_1=0.35$, $M_{S-R}=0.30\mu\Phi_0/$nA; for $\beta_1=0$, $M_{S-R}=0.27\mu\Phi_0/$nA. The kink in the blue line for sample 13 is an artifact of the algorithm that interpolates between our calculated points.} 
\end{figure}

For most of the samples, the best fit was obtained when a uniform cross section was assumed. However, there were a few samples for which a better fit was obtained for $\beta_1\neq 0$, and two of them are presented in Fig.~\ref{Nicdata}, each for a temperature above or close to $T_c-\TLP$ and another temperature well below $T_c-\TLP$. The temperature of P$_2$ is $\sim T_c-(1+2\beta_1)\TLP$,\cite{book,SIAM} so that we may expect nonuniformity to extend the range over which there is no hysteresis at $\Phi=\Phi_0/2$ by an amount of $\sim 2\beta_1\TLP$.
In Fig.~\ref{Nicdata}, the lines marked with a temperature are the experimental representative curves for $I(\Phi)$, the points are calculated values using our method, and the lowest line is an interpolation for values calculated assuming a uniform cross section. Sample 13 is the one with largest deviation from uniformity and, indeed, SEM inspection reveals that this ring has imperfections. In general our calculated points are in good agreement with the experimental lines. For the lower temperatures shown in Fig.~\ref{Nicdata}, the theoretical curves are more rounded than the experimental curves; this is probably due to our model assumption that the cross section profile is sinusoidal. 

\subsection{Wide Rings}
\subsubsection{Current--flux curves\label{CFC}}
In this section we deal with the samples of Ref.~\onlinecite{Nick} that have $D>R/4$.
As in the case of 1D rings, we generated representative curves $I(\Phi)$ for the experimental data, but slightly different strategies were adopted: the overall slope of the data was not subtracted, the shift in the origin to render $I(\Phi)$ an odd function was fixed ``by eye," and different polynomia were used to fit the data in different ranges.

As discussed in Sec.~\ref{widesecII}, the contribution of the induced magnetic field is not quite negligible. Since the field generated by self induction is not uniform, and since additional magnetic induction mechanisms may be present besides self induction, we initially attempted to regard $L_s$ as an adjustable parameter. The results we obtained were of the order of the tabulated values,\cite{tables} but systematically smaller; in some cases we obtained $L_s<0$. In view of this behavior, and in order to reduce the number of parameters, we decided to neglect self induction altogether. This omission may influence the effective value of $\beta_1$.

\begin{figure}
\scalebox{0.85}{\includegraphics{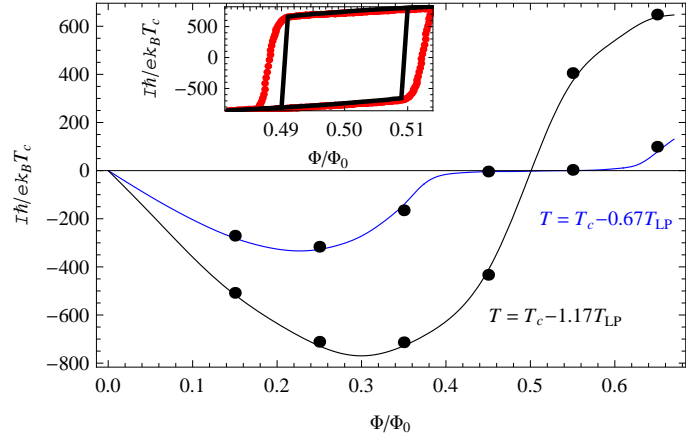}}%
\caption{\label{s5}Current as a function of the magnetic flux for a wide sample. The lines are smoothed experimental data and the dots were calculated.
The larger currents are for $T=T_c-1.17T_{\rm LP}$ and the smaller currents (blue online) for $T=T_c-0.67T_{\rm LP}$. Parameters taken from Ref.~\onlinecite{Nick}: $L=2.2\mu$m, $\bar w=7.83\times 10^{-11}{\rm cm}^2$, $D=1.35\times 10^{-5}{\rm cm}$, $T_c=1.240$K; parameters fitted here: $\ell=30.2$nm, $\beta_1=0.07$ and mutual inductance $M_{S-R}=0.035\mu\Phi_0/$nA. Inset: hysteresis region; the curves (actually, dense set of points, red online) are experimental data for $T=T_c-1.56T_{\rm LP}$ and the black lines are calculated.} 
\end{figure}

Figure \ref{s5} shows our results for sample 5. This sample had a width/radius ratio $D/R=0.39$. Note that for this sample the vOR method could not be applied.\cite{Nick} We chose the value of $\ell$ from the data for $T=T_c-0.17T_{\rm LP}$, the value of $\beta_1$ was adjusted to reproduce the experimental ratio $I(0.507\Phi_0)/I(0.1\Phi_0)$ at $T=T_c-1.56T_{\rm LP}$ and the mutual inductance was adjusted at $T=T_c-1.17T_{\rm LP}$. We also used the data at $T=T_c-1.56T_{\rm LP}$ to refine the calibration of the applied flux. Our calculated results in the hysteresis region (inset) are in surprising agreement with the experiment. However, the calculated decays of metastable states are much sharper than the observed decays; the lack of sharpness in the observed decay is probably due to the fact that the experiment was not perfectly static: the range of flux involved in the passage between fluxoid states was swept during a lapse of time of the order of $10^{-4}\,$sec, which is comparable with the filtering time.

\begin{figure}
\scalebox{0.85}{\includegraphics{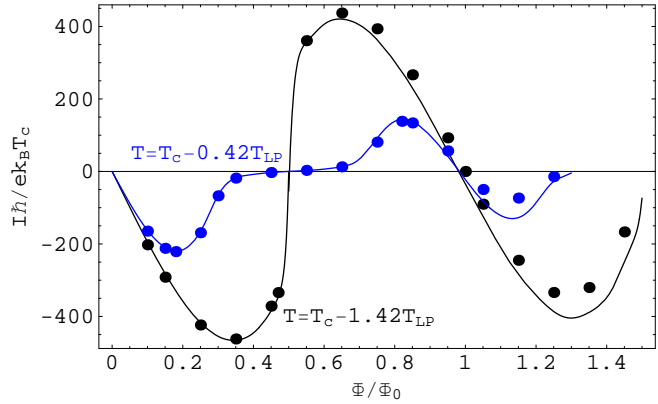}}%
\caption{\label{s12}Current as a function of the magnetic flux for another wide sample. The larger currents are for $T=T_c-1.42\TLP$ and the smaller currents (blue online) for $T=T_c-0.42\TLP$. For visibility, the currents in the case $T=T_c-0.42\TLP$ have been multiplied by the factor 3. Parameters taken from Ref.~\onlinecite{Nick}: $L=3.14\mu$m, $\bar w=1.01\times 10^{-10}{\rm cm}^2$, $D=1.75\times 10^{-5}{\rm cm}$, $T_c=1.244$K; parameters fitted here: $\ell=30.9$nm, $\beta_1=0$ and mutual inductance $M_{S-R}=0.082\mu\Phi_0/$nA.} 
\end{figure}

Figure \ref{s12} shows our results for sample 12, which had a width/radius ratio $D/R=0.35$. Among the samples that we consider as wide, this is the one for which the largest flux was reached in the experiment. The value of $\ell$ was adjusted using the data for $T=T_c-0.42\TLP$ and $\beta_1$ and the mutual inductance were adjusted using the data for $T=T_c-1.42\TLP$. We see that there is fair agreement between our model and the experimental results for $\Phi\alt\Phi_0$, but for larger fluxes there is a considerable deviation. The experimental currents vanish for $\Phi\approx 0.98\Phi_0$, whereas the calculated currents vanish almost exactly at $\Phi=\Phi_0$. 

\begin{figure}
\scalebox{0.85}{\includegraphics{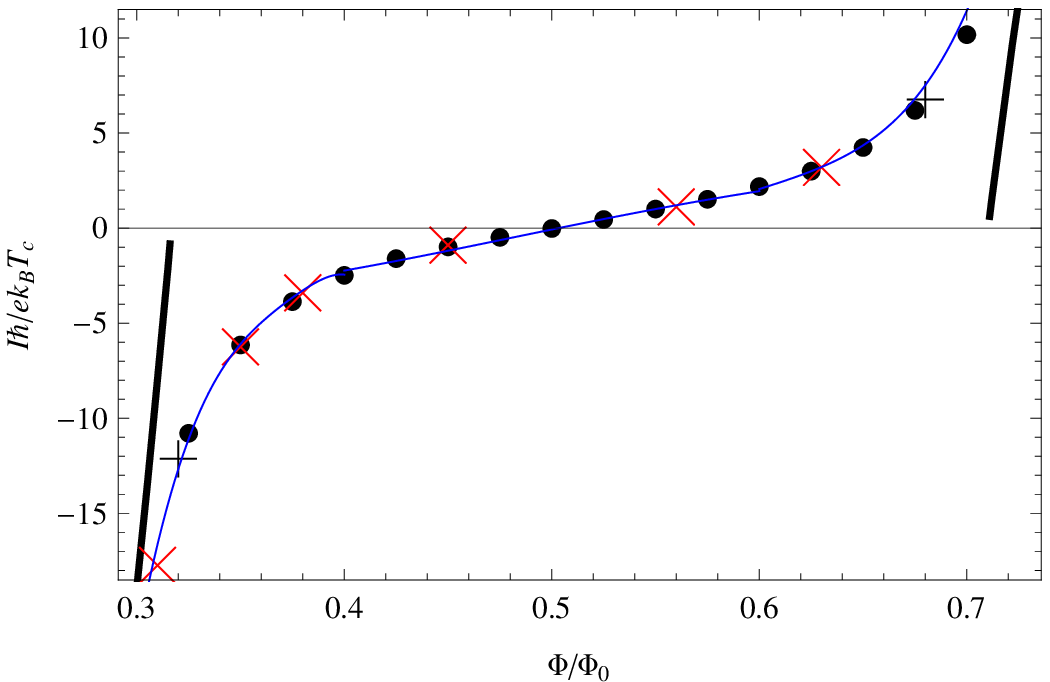}}%
\caption{\label{flr}Blowup of the fluctuation region around $\Phi=0.5\Phi_0$ for $T=T_c-0.42\TLP$ in Fig.~\ref{s12}. The oblique lines of straight appearance at the left and at the right describe the currents that would be obtained without thermal fluctuations. The curve describes the smoothed experimental results and the symbols were calculated. $\bullet$ TDGL; $\times$ KWT (Sec. \ref{kwt}), $\tau_{\rm ph}=10^{-9}\,$sec; + KWT, $\tau_{\rm ph}=10^{-8}\,$sec.} 
\end{figure}

The fluctuation region around $\Phi=0.5\Phi_0$ for sample 12 is shown enlarged in Fig~\ref{flr}. In the absence of fluctuations, the current would vanish for $0.32\le\Phi/\Phi_0\le 0.71$. The lines that show the current that would be present without fluctuations were obtained with the same code as all the other results, but the Langevin terms were divided by a factor of 100. Note that all the adjustable parameters of our model were fixed outside this region.

Our fitted values of $\ell$ vary from sample to sample. The largest value equals almost twice the smallest value, but they are all (including thin and wide samples) within 15\% of the empirical expression $1/\ell =1.5\times 10^5 {\rm cm}^{-1}+2/D+14.5/L$. Our values of $\ell$ are larger than those of Ref.~\onlinecite{Nick}; the difference is mainly due to the adopted value of $v_F$.

\subsubsection{Flux-induced vortices}
A recent experiment in a doubly connected asymmetric disk\cite{Kanda,JB} found that, for the examined sample, the order parameter practically vanishes in the narrow part of the sample when passing between the fluxoid state 0 and the fluxoid state 1; when passing between 2 and 3 the order parameter practically vanishes in the wide part, and no local vanishing is observed in the passage between 1 and 2.

We consider now a sample with geometric and material parameters similar to those reported in the experiment and check whether our theoretical method reproduces the experimental behavior. Our results are shown in Table \ref{vortex}. We denote by $\psi_\theta$ the order parameter at angle $\theta$, where $\theta=0$ is the angle at which the sample is widest; $\langle\cdots\rangle$ is average over time; for each temperature and flux we use the normalization constant $C=0.5(\langle |\psi_{\pi/2}|^2\rangle+\langle |\psi_{-\pi/2}|^2\rangle)$; stdev$_\theta$ stands for $(\langle |\psi_\theta|^4\rangle -\langle |\psi_\theta|^2\rangle ^2)^{1/2}$. Indeed we see that for $\Phi=0.5\Phi_0$ and for a temperature within the appropriate experimental range $\langle |\psi|^2\rangle$ is particularly small for $\theta=\pi$, whereas for $\Phi=2.5\Phi_0$ the small value is obtained for $\theta=0$. In the table we show the first two moments only; higher moments indicate that in these two situations $|\psi|^2$ has exponential distribution.

\begin{table}
\caption{\label{vortex}Angular dependence of the characteristic size of the order parameter for a sample with radius 
290\,nm, $D(\theta )=168(1+0.625\cos\theta )$\,nm, $\bar w=5.04\times 10^{-11}{\rm cm}^2$, $T_c=1.36$K and $\ell=15.6$\,nm. $C=0.5(\langle |\psi_{\pi/2}|^2\rangle+\langle |\psi_{-\pi/2}|^2\rangle)$. Appropriate temperatures were taken fron Ref.~\onlinecite{Kanda}}
\begin{ruledtabular}
\begin{tabular}{cccccc}
% Lines of table here ending with \\
$\Phi/\Phi_0$ & $T$(K) & $\langle |\psi_0|^2\rangle/C$ & stdev$_0/C$ &  $\langle |\psi_\pi|^2\rangle/C$ & stdev$_\pi/C$ \\
\hline
0.5 & 1.23 & 1.23 & 0.08 & 0.016 & 0.017 \\
1.5 & 1.20 & 0.64 & 0.61 & 1.23 & 1.19 \\
2.5 & 1.05 & 0.012 & 0.012 & 4.45 &  0.79
\end{tabular}
\end{ruledtabular}
\end{table}

\begin{figure}
\scalebox{0.85}{\includegraphics{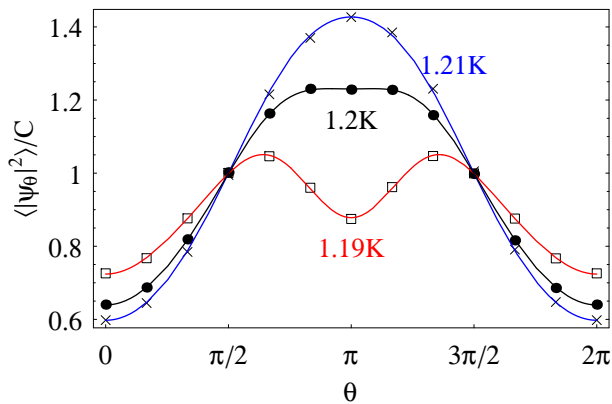}}%
\caption{\label{psqtet}Average value of $|\psi|^2$ as a function of the position in the sample for $\Phi=1.5\Phi_0$. The sample is widest at $\theta =0$ and narrowest at $\theta =\pi$. The sample properties are described in Table \ref{vortex}. $\Box$ ($\bullet$, $\times$) $T=1.19$\,K ($T=1.2$\,K, $T=1.21$\,K), 12 computational cells. Typically, we took $3\times 10^8$ averaging steps, $6\times 10^7$ relaxation steps, each step of $3\times 10^{-6}$ time units. The lines are guides for the eye. Note that the normalization $C$ depends on the temperature.}
\end{figure}

Moreover, for $\Phi=1.5\Phi_0$, $\langle |\psi_\theta|^2\rangle$ is not small for any $\theta $, in agreement with the experimental observation. Figure \ref{psqtet} shows that for this flux and temperature $\langle |\psi_\theta|^2\rangle$ does not depend very strongly on $\theta $. However, we note that stdev$_0$ (respectively stdev$_\pi$) is almost as large as $\langle |\psi_0|^2\rangle$ ($\langle |\psi_\pi|^2\rangle$), suggesting that there are frequent transitions in both directions between the fluxoid states 1 and 2. Since a continuous change in winding number must involve a place where the order parameter vanishes, we may anticipate that small averages of $|\psi|^2$ will be obtained if we average over transition steps only.

For this purpose we need a criterion to decide what is meant by a ``transition step." We start by noting that in the fluxoid state 2 a positive current flows around the sample for $\Phi\approx 1.5\Phi_0$, whereas in the state 1 the current is negative, so that in a transition between the two states the current should change sign. However, there may be many small steps in which the sign changes back and forth; we will not count each of these changes as a transition, but only the first one after the sample has been in a ``typical" state. Let us denote by $I_+$ ($I_-$) the average of the current over those steps in which it is positive (negative). We found that $|I_-|\approx I_+$. We regarded a step as a ``transition up" (down) if it is the first one for which the current is positive (negative) after having been smaller than $-I_+/8$ (larger than $I_+/8$). When a transition step is detected, we identify the cell in which $|\psi|^2$ has its minimum value as the place where a phase slip occurs.

\begin{figure}
\scalebox{0.85}{\includegraphics{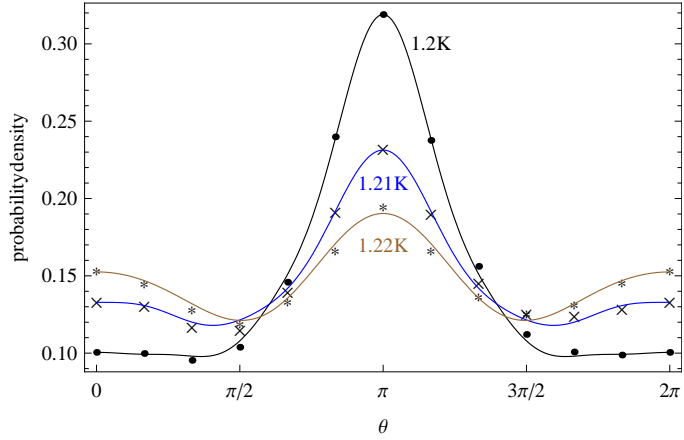}}%
\caption{\label{density}Probability per unit angle for the location of the vortex when the fluxoid state of the sample switches between 1 and 2. * $T=1.22$\,K; other parameters and symbols as in Fig.~\ref{psqtet}.} 
\end{figure}

Figure \ref{density} shows the probability density for the location of the phase slip in a fluxoid transition as a function of the angle $\theta$, for $\Phi=1.5\Phi_0$ and for several temperatures close to the onset of superconductivity. There is no appreciable difference between the probability distributions for transitions up or transitions down, and therefore we present the distributions for transitions in both directions. The probability distributions do not depend appreciably on the criterion for ``typical state"; for instance, using the threshold $I_+/4$ rather than $I_+/8$ leads to practically the same distribution. The probability distributions slightly depend on the value of the conductivity; the effect of raising the value of $\sigma $ is similar to that of lowering the temperature. The results in Fig.~\ref{density} were obtained for $\sigma =1.22\times 10^{11}\ell\Omega^{-1}{\rm cm}^{-2}$.

As expected, transition steps occur when there is at least one cell where $|\psi|^2$ is small. If we denote by ${\cal A}$ the event that there is a cell where $|\psi|^2<C/20$ and by ${\cal B}$ the event that there is a transition step, then, for $T=1.2\,$K and $\Phi=1.5\Phi_0$, the probability of having simultaneously ${\cal A}$ and ${\cal B}$ is 1.9 times larger than the product of the probabilities of ${\cal A}$ and ${\cal B}$. Transition steps are clustered: denoting by $t_{\rm tr}$ the time interval between consecutive transition steps and by overline the average over intervals, the relative standard deviation is $(\overline{t_{\rm tr}^2}-\overline{t_{\rm tr}}^2)^{1/2}/\overline{t_{\rm tr}}=3.7$. This result is consistent with a scenario in which most of the time the order parameter is too large to permit transitions and, every time it becomes suficiently small, several transitions occur.

\subsubsection{Life-time of metastable states\label{life-}}

Zhang and Price\cite{ZP} performed direct measurements of the lifetime of metastable states, as a function of the temperature and the flux. They prepared states with winding number 0 that enclosed flux larger than $0.5\Phi_0$ and waited until a change in the magnetic susceptibility was detected. The elapsed time was measured and the same procedure was repeated several times, until a significant average was obtained.

There are many reasons not to expect that our method will be able to reproduce these experimental values of the lifetime. First, TDGL is not expected to be valid at the considered temperatures; second, the width/radius ratio for their sample was 0.54, so that the quasi-1D description seems inappropriate; third, the direct measurements could be performed for lifetimes larger than $1\,$sec, whereas the lifetimes that we could practically study were smaller than $10^{-3}\,$sec; fourth, since the fluctuations above $T_c$ in Ref.~\onlinecite{ZP} were much larger than predicted by theory, we may suspect that some non-thermal perturbation was present in the experiment and that this perturbation (rather than thermal activation) could induce the decays. In spite of all these difficulties, having an estimate of the influence of nonuniformity on the lifetime of a metastable state is at least of academic interest.

\begin{table}
\caption{\label{life}Flux for which the lifetime of the metastable state is $10^{-4}\,$sec. The values of $\phi_{-4}$ in the 2$^{\rm nd}$ column are experimental and in the 3$^{\rm rd}$ and 4$^{\rm th}$ column were calculated using TDGL; the last two columns are for the Kramer--Watts-Tobin model (Sec.~\ref{kwt}).
Physical parameters: 
radius=950\,nm, $D=515$\,nm, $\bar w=1.03\times 10^{-10}{\rm cm}^2$, $T_c=1.266\,$K, $\sigma =3.88\times 10^5\Omega ^{-1}{\rm cm}^{-1}$, $\ell=31.8$\,nm, $\tau_{\rm ph}=10^{-8}$\,sec. $N=5$; the time step was $3\times 10^{-4}$ time units.}
\begin{ruledtabular}
\begin{tabular}{cccccc}
% Lines of table here ending with \\
$T$(K) & $\phi_{-4}$ (exp.) & $\phi_{-4}^{\rm TDGL}$ ($\beta_1=0$) & $\phi_{-4}^{\rm TDGL}$ ($\beta_1=0.25$) &  $\phi_{-4}^{\rm KWT}$ ($\beta_1=0$) & $\phi_{-4}^{\rm KWT}$ ($\beta_1=0.25$)\\
\hline
1.192 & 0.63 & 0.72  & 0.62 & 0.79 & 0.62\\
1.137 & 0.79 & 0.88 & 0.73 & &
\end{tabular}
\end{ruledtabular}
\end{table}

Let us denote by $\phi_{-4}$ the value of $\Phi_x/\Phi_0$ for which the lifetime of the state with winding number 0 is $10^{-4}\,$sec. The values of $\phi_{-4}$, for temperatures at which measurements were performed and are not too far from $T_c$, are shown in Table \ref{life}. The experimental value of $\phi_{-4}$ was determined by extrapolating the exponential dependence found in Ref.~\onlinecite{ZP}. For comparison, $\phi_{-4}$ was also calculated with our method, using the physical parameters reported as the best fit in Ref.~\onlinecite{ZP} and two cross section profiles; in one case the cross section was taken as uniform and in the other case we assumed uniform width, whereas the thickness had sinusoidal dependence, with $\beta _1=0.25$, which is not far from the value 0.15 used for the same sample in Ref.~\onlinecite{book} in order to mimic the shape of the phase diagram. The initial values of the order parameter were those that minimize the free energy for winding number 0; evolution followed governed by TDGL with fluctuations, and we decided that the initial state had decayed when a positive current was reached; after a decay, the process was repeated, during a total of $3\times 10^5$ time units.

Although in Sec. \ref{CFC} we ignored self inductance, in the present case flux sensitivity is very large and we therefore decided to take self inductance into account. Self inductance was taken into account also in Ref.~\onlinecite{ZP}, but the influence they found had opposite sign than what we find here: in our (resp. their) case self inductance increases (resp. reduces) metastability. In our case the main effect of self inductance is due to the difference between $\Phi_x$ and $\Phi$; in their case the self inductance is used in a model for estimating the height of the potential barrier.

Calculated lifetimes divided by their averages have typically Poissonic distributions.

While bearing in mind the reservations raised at the begining of this section, Table \ref{life} suggests that nonuniformity could be a reason for the low lifetimes found in Ref.~\onlinecite{ZP}.

\section{\label{kwt}Beyond TDGL}
\subsection{The Kramer--Watts-Tobin model}
Kramer and Watts-Tobin\cite{KWT} (KWT) extended TDGL so as to render it applicable to gapped superconductors and valid as long as there is local equilibrium. The KWT model has been successfully used to describe the current-voltage characteristic of thin wires.\cite{Ivlev,Tidecks,S-shape}
The energy functional is the same as in the Ginzburg--Landau model; in a gauge such that the electrochemical potential is uniform, Eq.~(\ref{macu}) generalizes to 
\be
\frac{1}{\sqrt{1+K|\psi_k|^2}}\left[ \frac{d}{dt}+\frac{K}{2}\frac{d|\psi_k|^2}{dt}\right] {\rm Re}[\psi_k]=-\Gamma_{\psi,k} \frac{\partial G}{\partial {\rm Re}[\psi_k]} \; 
\label{macuKWT}
\ee
and similarly for ${\rm Im}[\psi_k]$. Here $K=247\kappa^2e^2k_BT_c\ell v_F\tau_{\rm ph}^2/\hbar mc^2$, where $\tau_{\rm ph}$ is the electron-phonon inelastic scattering time. In the limit $K\rightarrow 0$, TDGL is recovered.
Aluminum is challenging since it has a particularly large $\tau_{\rm ph}$, i.e., $\tau_{\rm ph}=10^{-8}$\,sec. As a consequence, the expected correction to TDGL should be large. Also, the range of validity of KWT should be small, $T_c-T\alt\hbar /k_B\tau_{\rm ph}\sim 1$\,mK.

Equation (\ref{macuKWT}) can be brought to canonic form by writing $\psi_k=|\psi_k |\exp (i\chi_k)$. We obtain
\begin{eqnarray}
d|\psi_k|/dt&=&-h_{|\psi|}(\psi_k)\Gamma_{\psi,k}\partial G/\partial |\psi_k| \;, \label{WTr} \\
d\chi_k/dt&=&-h_\chi(\psi_k)\Gamma_{\psi,k}\partial G/\partial \chi_k \;,
\label{WT}
\end{eqnarray}
with $h_{|\psi|}(\psi_k)=(1+K|\psi_k|^2)^{-1/2}$ and $h_\chi(\psi_k)=1/(h_{|\psi|}|\psi_k|^2)$.

\subsection{Appropriate Langevin terms}

Following the reasoning of Ref.~\onlinecite{Langevin}, for a period of time $\tau$ over which $|\psi_k|$ does not change significantly, the fluctuating parts of the changes of $|\psi_k|$ and of $\chi_k$ would be expected to have variances $\langle \eta_{|\psi|}^2\rangle =2h_{|\psi|}(\psi_k)\Gamma_{\psi,k} k_BT\tau$ and $\langle \eta_\chi^2\rangle =2\Gamma_{\psi,k} k_BT\tau/h_{|\psi|}(\psi_k)|\psi_k|^2$. However, we have pointed out elsewhere\cite{balance} that in the case of Eq.~(\ref{WTr}) this is not the full story. Due to the $|\psi_k|$-dependence of $h_{|\psi|}$ and of the Jacobian $W=\partial ({\rm Re}[\psi_k],{\rm Im}[\psi_k])/\partial (|\psi_k|,\chi_k)$, fluctuations of $|\psi_k|$ do not have zero average but rather $\langle \eta_{|\psi|}\rangle =[\partial \log(h_{|\psi|}W)/\partial |\psi_k|]h_{|\psi|}\Gamma_{\psi,k} k_BT\tau$. Using the expressions for $h_{|\psi|}$ and $W$ we obtain
\be
|\psi_k|(t+\tau)-|\psi_k|(t)=h_{|\psi|}\Gamma_{\psi,k}\left[\frac{h_{|\psi|}^2 k_BT}{|\psi_k|}-\frac{\partial G}{\partial |\psi_k|}\right]\tau+\bar\eta_{|\psi|}\;,
\label{polar}
\ee
where $\bar\eta_{|\psi|}$ has gaussian distribution, $\langle \bar\eta_{|\psi|}\rangle =0$ and $\langle \bar\eta_{|\psi|}^2\rangle =2h_{|\psi|}(\psi_k)\Gamma_{\psi,k} k_BT\tau$.

\subsection{Results and comparison with experiments}
We have limited our study to two experiments. The first is the case of persistent current in a wide ring in the fluctuation region, shown in Fig.~\ref{flr}. The oblique crosses are for $\tau_{\rm ph}=10^{-9}\,$sec and the upright crosses for $\tau_{\rm ph}=10^{-8}\,$sec; the other parameters are as for TDGL. We see that even for large values of $\tau_{\rm ph}$ the currents obtained coincide within statistical uncertainty with those obtained with TDGL.

\begin{figure}
\scalebox{0.85}{\includegraphics{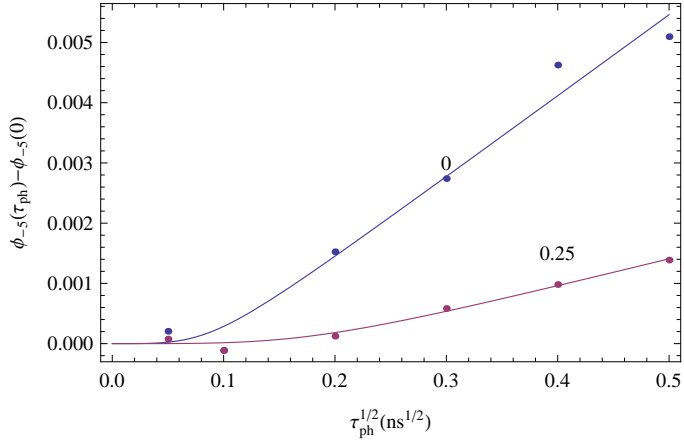}}%
\caption{\label{tauph}Increment of the flux for which the lifetime of a metastable state is 10$^{-5}\,$sec for $T=1.192\,$K, $\beta_1=0$ (blue online) and $\beta_1=0.25$ (violet online), for $0\le\tau_{\rm ph}\le 2.5\times 10^{-10}\,$sec. The other parameters are as in Table \ref{life}. The dots were calculated and the lines are empiric fits.}
\end{figure}

The second experiment we studied was that of the lifetime of metastable states, considered in Sec.~\ref{life-}. Figure \ref{tauph} shows our results for the flux at which the lifetime is $10^{-5}\,$sec. The results are reasonably fitted by expressions of the form
$\phi_{-5}(\tau_{\rm ph})=\phi_{-5}(0)+a[(b^2+\tau_{\rm ph}^2)^{1/4}-\sqrt{b}]$, where $a$ and $b$ are adjustable constants. Using a fit of this form for $\phi_{-4}$ and extrapolating to $\tau_{\rm ph}=10^{-8}\,$sec leads to the results shown in Table \ref{life} (The increment of $\phi_{-4}$ due to KWT for $\beta_1=0.25$ is less than 0.01).

\section{Discussion\label{Disc}}
We have evaluated the influence of thermal fluctuations on persistent currents (i.e., average current), the position of the flux-induced vortices, and lifetime of metastable fluxoid states for aluminum rings that enclose magnetic flux, in a range down to $\sim 10^2\,$mK below the critical temperature, using TDGL with the addition of Langevin terms. In the cases that we considered to be most interesting, the evaluation was also performed using the KWT model. The considered rings were not ``ideal," in the sense that they were either wide, so that the ``enclosed" flux was not sharply defined, or nonuniform, so that expansion of the order parameter into a Fourier series does not simplify the problem.
In all cases we obtained at least qualitative agreement between TDGL and the experimental results, and in many cases we obtained quantitative agreement, in spite of the fact that we considered temperatures far beyond the range where TDGL is justified by microscopic theory.

Nonuniformity leads to smaller persistent currents (in comparison with a uniform ring with the same average parameters), to smaller slopes of the current as a function of the flux when passing between fluxoid states, and to a larger temperature range for which the passage between fluxoid states is non-hysteretic. It also leads to a narrower range of metastability. Finite width leads to smaller persistent currents (in comparison with 1D rings) and to lesser sensitivity to nonuniformity.

For several experiments in which the rings were intended to be uniform, better agreement with theory can be obtained by assuming that the rings actually had some unintentional nonuniformity; for those rings that were intentionally nonuniform, good agreement with theory was found using the reported values for the shape of the ring.

Several years ago it was predicted that the passage between fluxoid states may be mediated by a flux-induced vortex\cite{fivortex} and this prediction has been experimentally confirmed.\cite{Kanda,JB} Our present study confirms the existence and position of this vortex. Moreover, it describes a situation in which the intermediate state between two fluxoid states is actually a dynamic situation with sporadic migrations between both states, involving a place where the order parameter vanishes.

The Kramer--Watts-Tobin (KWT) model was applied to two situations, mainly in order to estimate the expected discrepancy incurred when using TDGL. The first case was that of the persistent current in a fluctuation range, i.e., a range for which the current would vanish in the absence of thermal fluctuations. Remarkably, even for the large value of $\tau_{\rm ph}$ in the case of aluminum, the currents obtained for KWT agree with those obtained for TDGL, and they both agree with the measured values. This agreement may be understood if we bear in mind that the persistent current is an  equilibrium quantity and therefore depends only on the energy spectrum and not on the dynamics; since the energy functional for KWT is the same as for TDGL, they both lead to the same currents. In more general terms, we can say that even if some features of a model, such as TDGL, are not an accurate description of reality, the model can accurately describe situations in which these features are irrelevant.

The second case in which we used KWT was the study of the lifetime of metastable states. We find that the ring can remain in the metastable state for longer times than predicted by TDGL. Since according to Eqs.~(\ref{WTr}) and (\ref{WT}) KWT brings about a slower evolution of the absolute value of the order parameter and a faster evolution of its phase, we can conclude that the evolution rate of the absolute value is dominant in determining metastability. This can be understood, since the winding number of the order parameter has to change in order to escape from a fluxoid state and this cannot happen as long as the absolute value of the order parameter is positive everywhere; on the other hand, when the order parameter does vanish at some point, the fast evolution of the phase could lead to multiple transitions, as discussed in Ref.~\onlinecite{oldjumps}.

\begin{acknowledgments}
I have benefited from correspondence with Konstantin Arutyunov, Ora Entin-Wohlman, Yuval Oreg, Hamzeh Roumani and Andrei Zaikin. I am especially grateful to Hendrik Bluhm for many useful remarks and to Nicholas Koshnick for sending me the entire data file of the experiment in Ref.~\onlinecite{Nick} and for explaining to me several unpublished details.
\end{acknowledgments}

\appendix
\section{\label{AA}Numerical aspects of KWT}
In the case of TDGL, we may regard the evolution of the order parameter as isotropic in $\psi$-space. On the other hand, Eqs.~(\ref{WTr}) and (\ref{WT}) show that according to KWT variations of $\psi_k$ in the ``radial" direction are inhibited by a factor $h_{|\psi|}$, whereas in the ``angular" direction they are  enhanced by a factor $1/h_{|\psi|}$. 
For situations in which $K|\psi_k|^2\gg 1$, this factor can be important, and this forces us to divide steps in the angular direction into several steps, i.e., for every step lasting time $\tau$ in the radial direction there are $n_\chi$ steps in the angular direction, each lasting time $\tau/n_\chi$ ($n_\chi\sim h_{|\psi|}^{-2}$). 

In cases in which we evaluate an equilibrium property, it might be unnecessary to perform all the steps in the angular direction, since, due to the fast variation of $\chi_k$ while $|\psi_k|$ remains frozen, we may assume that after a relatively small number of steps $n_{\rm prob}\ll n_\chi$ we have already obtained the equilibrium value of the evaluated property for the given value of $|\psi_k|$, and we may go on and probe new values of $|\psi_k|$. In this case it is important to note that the statistical weight of this partial average is $\tau$ and not $n_{\rm prob}\tau /n_\chi$.

Equation (\ref{polar}) was implemented through the assignment $\psi_k(t+\tau)=\psi_k(t)(1+\Delta |\psi_k|/|\psi_k|)$. Some caution is required due to the presence of the factor $|\psi_k|^2$ in the denominator, which might accidentally be very small. We replaced it with $|\psi_k|^2+\tau\bar\psi^2/\bar\tau$; unless $|\psi_k|$ is exceptionally small, this replacement gives rise to a negligible $O(\tau^2)$ contribution, while it guarantees that $\psi_k$ remains conveniently bounded in these exceptional cases.

\section{\label{BB}Persistent Currents in Normal Rings}
\begin{figure}
\scalebox{0.85}{\includegraphics{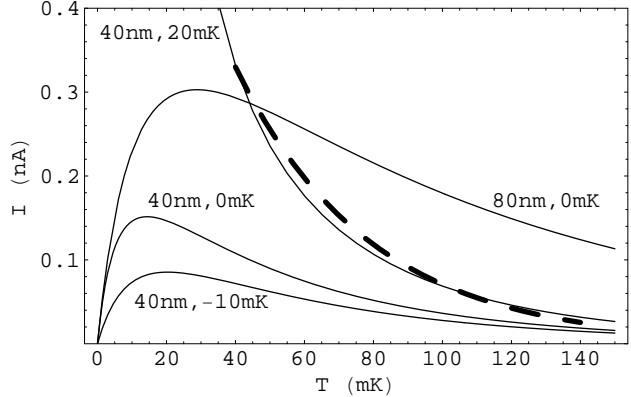}}%
\caption{\label{myfig}Size of the current as a function of temperature for $\phi=0.25$. The continuous lines were calculated using Eq.~(\ref{final}) and the values of $\ell$ and $T_c$ are marked next to each line; the other parameters were taken from Ref.~\onlinecite{NM3}. The mean free path obtained from normal conductivity measurements is $\ell =40$\,nm. The dashed line is the fit to the measurements given in Ref.~\onlinecite{NM3}. } 
\end{figure}

In normal metals current is due to single electrons, so that persistent currents are expected to have a $2\Phi_0$ flux period; the size and sign of the current are governed by $\cos(k_FL)$, which is random in practice. Recent experiments\cite{Bluhm,Science} on single rings are consistent with theoretical predictions.

On the other hand, several experiments on arrays of isolated rings of normal metals\cite{NM1,NM2,NM3} exhibit periodicity, dependence on the number of rings, sign and orders of magnitude that would be appropriate for superconductors. 
Available theories\cite{AE,Imry,ImryArx} indeed regard the average persistent currents as fluctuation superconductivity.

Since one of the objectives of this article is to check the applicability of the Ginzburg--Landau model to situations that are not guaranteed by a microscopic treatment, we compare this model with one of the experiments. By using this model, we are considering thermal fluctuations only, whereas quantum fluctuations are being ignored. According to Ref.~\onlinecite{ImryArx}, this approach should give the dominant contribution to the current for temperatures $\agt\TLP$. The critical temperature $T_c$ will be regarded as a phenomenologic parameter, such that $T_c>0$ describes a superconductor, $T_c=0$ describes a perfect normal conductor, and the possibility of an ``antisuperconductor" with $T_c<0$ (which could be due to repulsive interaction among the electrons) can also be contemplated within this model.

We consider a two-dimensional narrow uniform ring, write the Ginzburg--Landau energy in terms of the gauge-invariant order parameter, and neglect the quartic term. The energy becomes
\be
G=\frac{\hbar^2w}{2mD}\int_0^Lds\int_{0}^D dr\left(\frac{4\pi^2\eta}{L^2}|\ps|^2+|\nab\ps|^2\right) \;,
\ee
with $\eta =mL^2\alpha /2\pi^2\hbar ^2=0.19k_B(T-T_c)L^2/\hbar\ell v_F$. $\ps$ obeys the boundary conditions $\ps (s+L)=e^{2\pi i\phi}\ps (s)$, where $\phi=\Phi/\Phi_0$, and $\partial\ps/\partial r=0$ at $r=0$ and $r=D$.

Performing the transformation
\be
\ps(s,r)=\sum_{n=-\infty}^\infty e^{2\pi i(n+\phi)s/L}\left(\varphi_{n0}+\sum_{j\ge 1}\varphi_{n,j} \cos\frac{\pi jr}{D}\right)  \;,
\label{fourier}
\ee
the energy becomes
\be
G=\frac{2\pi^2\hbar ^2w}{mL}\sum_{n=-\infty}^\infty\left\{\left[\eta +(n+\phi)^2\right]\left[|\varphi_{n0}|^2+\frac{1}{2}\sum_{j\ge 1}|\varphi_{n,j}|^2\right]+\frac{L^2}{8D^2}\sum_{j\ge 1}j^2|\varphi_{n,j}|^2\right\} \;,
\ee
hence the statistical averages $\langle |\varphi_{n0}|^2\rangle =mLk_BT/2\pi^2\hbar ^2w[\eta +(n+\phi)^2]$ and $\langle |\varphi_{nj}|^2\rangle =mLk_BT/\pi^2\hbar ^2w[\eta +(n+\phi)^2+(L^2j^2/4D^2)]$.

The persistent current can be written as $\langle I\rangle =-(2e\hbar w/mLD)\langle {\rm Im}\int_0^Lds\int_{0}^D dr\;\ps^*\partial \ps/\partial s\rangle$. Using transformation (\ref{fourier}) we obtain
\be
\langle I\rangle =-\frac{4\pi e\hbar w}{mL}\sum_{n=-\infty}^\infty (n+\phi)\langle |\varphi_{n0}|^2+\frac{1}{2}\sum_{j\ge 1}|\varphi_{n,j}|^2\rangle
\ee
and introducing the statistical averages the current becomes
\be
\langle I\rangle =-\frac{2 ek_BT}{\pi\hbar }\sum_{n=-\infty}^\infty (n+\phi)\sum_{j=0}^{\infty}\left[\eta +(n+\phi)^2+\frac{j^2L^2}{4D^2}\right]^{-1}\;.
\ee
Summation over $j$ gives
\be
\langle I\rangle =-\frac{ek_BT}{\pi\hbar }\sum_{n=-\infty}^\infty\frac{(n+\phi)[1+f(2\pi D\sqrt{\eta +(n+\phi)^2}/L)]}{\eta +(n+\phi)^2} \;,
\label{almost}
\ee
with $f(x)=x\coth x$; for the relevant experimental values, $f(x)$ may be replaced with 1, which is the zero-width limit. 

The series in Eq.~(\ref{almost}) is not absolutely convergent and we therefore need some physical criterion in order to sum it. Since $n+\phi$ is proportional to the velocity of Cooper pairs, and the sum should actually not contain the terms with velocities above the pair-breaking velocity, we introduce a cutoff for $|n+\phi|$. This is done by replacing Eq.~(\ref{almost}) with 
\be
\langle I\rangle =-\frac{ek_BT}{\pi\hbar }\sum_{n=-\infty}^\infty\frac{(n+\phi)[1+f(2\pi D\sqrt{\eta +(n+\phi)^2}/L)]}{\eta +(n+\phi)^2+\epsilon (n+\phi)^4} \;,
\label{final}
\ee
where $\epsilon $ is a small number. Taking $\epsilon =0.01$ or $\epsilon =0.001$, or adding $\epsilon (n+\phi)^6$ rather than $\epsilon (n+\phi)^4$, we obtain practically the same results.

As a function of the flux, Eq.~(\ref{final}) predicts a persistent current that vanishes when $\phi$ is an integer multiple of 0.5 and has maximum size when $\phi$ is close to an odd multiple of $0.25$. Figure~\ref{myfig} compares the size of the current at $\phi=0.25$ predicted by Eq.~(\ref{final}) with the experimental values obtained for silver rings. The values of the parameters were taken from Ref.~\onlinecite{NM3}. 
This experiment was not considered in Ref.~\onlinecite{ImryArx}, because measurements were performed at a high frequency.
In order to allow for the possibility that only a subset of the electrons on the Fermi surface are involved in superconductivity, we also considered values of the mean free path that are different from the average value deduced from normal conductivity. The best agreement with the experimental results is obtained for a rather large positive $T_c$, favoring the scenario advanced in Ref.~\onlinecite{Imry}.


\begin{thebibliography}{999}
\bibitem{LP}W. A. Little and R. D. Parks, Phys.\ Rev.\ Lett.\ {\bf 9}, 9 (1962).
\bibitem{AL} L. G. Aslamazov and A. I. Larkin, Fiz.\ Tver.\ Tela {\bf 10}, 1104 (1968) [Soviet Physics - Solid State {\bf 10}, 875 (1968)].
\bibitem{Oppen}F. von Oppen and E. K. Riedel, Phys.~Rev.~B {\bf 46}, 3203 (1992).
\bibitem{Buzdin}M. Daumens, C. Meyers, and A. Buzdin, Phys.~Lett.~A {\bf 248}, 445 (1998).
\bibitem{Oreg} G. Schwiete and Y. Oreg, Phys.\ Rev.\ Lett.\ {\bf 103}, 037001 (2009). 
\bibitem{ZP}X. Zhang and J.C. Price, Phys.\ Rev.\ B {\bf 55}, 3128 (1997).
\bibitem{Nick}N.C. Koshnick, H. Bluhm, M. E. Huber, and K. A. Moler, Science {\bf 318}, 1440 (2007).
\bibitem{Liu}M. M. Rosario, Yu. Zadorozhny, B.Y. Rock, P.T. Carrigan, H. Wang, and Y. Liu, Physica B {\bf 329-–333}, 1415 (2003).
\bibitem{McH}D.E. McCumber and B.I. Halperin, Phys. Rev. B {\bf 1}, 1054 (1970).
\bibitem{AV}A. Larkin and A. Varlamov, {\it Theory of Fluctuations in Superconductors} (Oxford University Press, Oxford, 2005).
\bibitem{Sch1}L. P. Gor'kov and G. M. Eliashberg, Zh. Eksp. Teor. Fiz. {\bf 54}, 612 (1968) [Soviet Phys. JETP {\bf 27}, 328 (1968)];
A. Schmid, Phys. Kondens. Mater. {\bf 5}, 302 (1966); M. Cyrot, Rep. Prog. Phys. {\bf 36}, 103 (1973).
\bibitem{AGZ}K.Yu. Arutyunov, D.S. Golubev and A.D. Zaikin, Physics Reports {\bf 464}, 1 (2008).
\bibitem{KWT} L. Kramer and R. J. Watts-Tobin, Phys. Rev. Lett. {\bf 40}, 1041 (1978); R. J. Watts-Tobin, Y. Kr\"{a}henb\"{u}hl and L. Kramer, J. Low Temp. Phys. {\bf 42}, 459 (1981).
\bibitem{Kanda}A. Kanda, B.J. Baelus, D.Y. Vodolazov, J. Berger, R. Furugen, Y. Ootuka, F.M. Peeters, Phys.\ Rev.\ B {\bf 76}, 094519 (2007).
\bibitem{JB}J. Berger, A. Kanda, R. Furugen, and Y.Ootuka, Physica C {\bf 468}, 848 (2008).
\bibitem{Langevin}J. Berger, Phys.~Rev.~B. {\bf 75}, 184522 (2007).
\bibitem{book}J. Berger in {\it Connectivity and Superconductivity} (J. Berger and J. Rubinstein, editors), Springer Verlag, Lecture Notes in Physics, vol. m62 (2000). 
\bibitem{JB0}J. Berger and J. Rubinstein, Phys. Rev. B {\bf 56}, 5124 (1997). 
\bibitem{Baelus} B. J. Baelus, F. M. Peeters, and V. A. Schweigert, Phys. Rev. B {\bf 61}, 9734 (2000).
\bibitem{jumps} D.Y. Vodolazov, F.M. Peeters, S.V. Dubonos, and A.K. Geim, Phys. Rev. B {\bf 67}, 054506 (2003).
\bibitem{SIAM} J. Berger and J. Rubinstein, SIAM J. Appl. Math. {\bf 58}, 103 (1998).
\bibitem{data} The entire set of currents as functions of sample, temperature and flux, was sent to us by the authors; the properties of each sample are tabulated in the Supporting Online Material for the article.
\bibitem{tables}F.W. Grover, {\it Inductance Calculations} (Dover, New York, 1973).
\bibitem{balance} arXiv:0909.0579v1 
\bibitem{Ivlev}B. I. Ivlev and N. B. Kopnin, Adv. Phys. {\bf 33}, 47 (1984).
\bibitem{Tidecks}R. Tidecks, {\it Current-Induced Nonequilibrium Phenomena in Quasi-One-Dimensional Superconductors} (Springer, Berlin,
1990).
\bibitem{S-shape} D.Y. Vodolazov, F.M. Peeters, L. Piraux, S. M\'{a}t\'{e}fi-Tempfli, and S. Michotte, Phys. Rev. Lett. {\bf 91}, 157001 (2003); S. Michotte, S. M\'{a}t\'{e}fi-Tempfli, L. Piraux, D. Y. Vodolazov, and F. M. Peeters, Phys. Rev. B {\bf 69}, 094512 (2004).
\bibitem{fivortex} J. Berger and J. Rubinstein, Phys.\ Rev.\ B {\bf 59}, 8896 (1999); D. Y. Vodolazov, B. J. Baelus, and F. M. Peeters, Phys.\ Rev.\ B
{\bf 66}, 054531 (2002).
\bibitem{oldjumps} D.Y. Vodolazov, F.M. Peeters, Phys.\ Rev.\ B {\bf 66}, 054537 (2002).

\bibitem{Bluhm} H. Bluhm, N. C. Koshnick, J. A. Bert, M. E. Huber, and K. A. Moler, Phys.\ Rev.\ Lett.\ {\bf 102}, 136802 (2009).
\bibitem{Science}A. C. Bleszynski-Jayich, W. E. Shanks, B. Peaudecerf, E. Ginossar, F. von Oppen, L. Glazman, and J. G. E. Harris, Science {\bf 326}, 272 (2009).
\bibitem{NM1}L. P. L\'{e}vy, G. Dolan, J. Dunsmuir, and H. Bouchiat, Phys.\ Rev.\ Lett.\ {\bf 64}, 2074 (1990).
\bibitem{NM2} E. M. Q. Jariwala, P. Mohanty, M. B. Ketchen, and R. A. Webb, Phys.\ Rev.\ Lett.\ {\bf 86}, 1594 (2001).
\bibitem{NM3} R. Deblock, R. Bel, B. Reulet, H. Bouchiat, and D. Mailly, Phys.\ Rev.\ Lett.\ {\bf 89}, 206803 (2002).
\bibitem{AE} V. Ambegaokar and U. Eckern, Europhys. Lett., {\bf 13}, 733 (1990).
\bibitem{Imry}H. Bary-Soroker, O. Entin-Wohlman, and Y. Imry, Phys.\ Rev.\ Lett.\ {\bf 101}, 057001 (2008).
\bibitem{ImryArx} H. Bary-Soroker, O. Entin-Wohlman, and Y. Imry, Phys.\ Rev.\ B {\bf 80}, 024509 (2009).

\end{thebibliography}
\end{document}